\documentclass[preprintnumbers,amsmath,amssymb,floatfix,11pt,prd,onecolumn,showpacs,superscriptaddress,nofootinbib]{revtex4}
\usepackage{graphicx}
\usepackage{epsfig}
\usepackage{bm}
\usepackage{amsfonts}
\usepackage{color}

\begin{document}

\title{Generalized Second Law of Thermodynamics in $f(T)$ Gravity with Entropy Corrections}
\author{\textbf{Kazuharu Bamba}}
\email{bamba@kmi.nagoya-u.ac.jp}
\affiliation{Kobayashi-Maskawa Institute for the Origin of Particles and the Universe, Nagoya
University,
Nagoya 464-8602, Japan}
\author{\textbf{ Mubasher Jamil}} \email{mjamil@camp.nust.edu.pk}
\affiliation{Center for Advanced Mathematics and Physics (CAMP),\\
National University of Sciences and Technology (NUST), H-12,
Islamabad, Pakistan}
\affiliation{Eurasian International Center
for Theoretical Physics, Eurasian National University, Astana
010008, Kazakhstan}
\author{\textbf{ D. Momeni}}
 \affiliation{Eurasian International Center
for Theoretical Physics, Eurasian National University, Astana
010008, Kazakhstan}
\author{\textbf{ Ratbay Myrzakulov}}
\email{rmyrzakulov@gmail.com}\affiliation{Eurasian International Center
for Theoretical Physics, Eurasian National University, Astana
010008, Kazakhstan}

\begin{abstract}

\textbf{Abstract:} 
We study the generalized second law (GSL) of thermodynamics 
in $f(T)$ cosmology. 
We consider the universe as a closed bounded system filled with $n$ component fluids in the thermal equilibrium with the cosmological boundary. We use two different cosmic horizons: the future event horizon and the apparent horizon. We show the conditions under which the GSL will be valid in specific scenarios of the quintessence and the phantom energy dominated eras. Further we associate two different entropies with the cosmological horizons: with a logarithmic correction term and a power-law correction term. We also find the conditions for the GSL to be satisfied or violated by imposing constraints on model parameters.
\end{abstract}
 \maketitle

\newpage

\section{Introduction}

A unification of quantum mechanics (QM) and general relativity (GR) is one of the goals of modern physics. Attempts to unify these two theories are termed Quantum Gravity in the literature. However a consistent theory of quantum gravity is not discovered yet, although several theories like loop quantum gravity and string theory predict new features about black holes and singularities. A well-known result is that general relativity predicts gravitational collapse of stars to black holes while quantum mechanics (together with GR) predicts their evaporation.  Both string theory and loop quantum gravity predict that a black hole
 emits thermal radiations whose thermal spectrum might deviate from the Planck black body spectrum at a certain small scale \cite{barton}, with a temperature proportional to its surface gravity at the black hole horizon and with an entropy proportional to its horizon area, the same predictions made by Hawking and Bekenstein several decades ago \cite{hawking}.  The Hawking temperature and horizon entropy together with the black hole mass obey the first law of thermodynamics. Some important discoveries connecting laws of thermodynamics and the Einstein field equations have been made 
\cite{paddy}. 
The Einstein equation has been derived from the Clausius relation in 
thermodynamics in GR~\cite{Jacobson:1995ab}, 
$f(R)$ gravity~\cite{Eling:2006aw, Elizalde:2008pv} and 
generalized gravitational theories~\cite{Brustein:2009hy}.
It has also been demonstrated that by applying the first law of thermodynamics to the apparent horizon of a 
Friedmann-Lema\^{i}tre-Robertson-Walker (FLRW) 
universe and assuming the geometric entropy given by a quarter of the apparent horizon area, the Friedmann equations of the universe with any spatial curvature can be derived \cite{cai}. Similar results also hold in scalar tensor, Gauss-Bonnet, Lovelock and $f(R)$ gravities \cite{akbar1} 
(for reviews on $f(R)$ gravity, see, e.g.,~\cite{Reviews-F(R)}). 
%
%

In this paper, we explore 
the GSL in the cosmological context and it can be interpreted that the time derivative generalized entropy (which has to be the sum of entropy of all fluids filling the space along with the entropy of the cosmological horizon) must be increasing (or non-decreasing) function of time. There have been some attempts to prove the GSL in different ways: a simple direct explicit proof of the 
GSL of black hole thermodynamics was proved for a quasistationary semiclassical black hole in Ref.~\cite{page}, also for two dimensional black hole spacetime \cite{shim}, using charged and rotating black holes \cite{wald}, by quantum information theory \cite{quantum}, using adiabatically collapsing thick light shells \cite{shells}. However the GSL may be violated in certain cases: It has been shown that classical non-minimally coupled scalar fields can violate all of the standard energy conditions and the GSL \cite{violate}. The GSL has found some interesting implications in string cosmology as well where it forbids singular string cosmologies \cite{string}.
In the study of accretion of phantom energy on black holes in cosmological background, the black holes will completely vanish if the GSL is violated \cite{rip}. In the framework of Horava-Lifshitz cosmology, it has been shown that under detailed balance the GSL 
is generally valid for flat and closed geometry and it is conditionally valid for an open universe, while beyond detailed balance the GSL is only conditionally valid for all curvatures \cite{hl}. In $F(R,G)$ gravity and modified gravity, the conditions for validity of the GSL have been discussed in \cite{sad} while in chameleon cosmology, 
the validity of the GSL in flat FRW chameleon cosmology where the boundary of the universe is the dynamical apparent horizon \cite{f} has been investigated. 
The GSL for fractional action cosmology by choosing various forms of scale factors was studied and proved that the GSL is not valid in all scenarios. For cosmological event horizon, it has been proved that the GSL is satisfied \cite{davies}. 
The conditions of validity of generalized second law in phantom energy dominated era has also been studied \cite{sad1}. However the GSL is violated when a black hole is introduced in a phantom energy dominated universe \cite{shit}. 
On the other hand, 
the GSL can be saved from the violation provided the temperature is not taken as de Sitter temperature \cite{sad2}. In a quintom (i.e. quintessence and phantom) dominated universe, the conditions for the GSL validation were 
investigated \cite{setare}.
For a holographic dark energy dominated universe, the GSL is respected for certain values of deceleration parameter \cite{setare1}. Moreover, the GSL has been studied in some lower-dimensional cosmological settings with 
several interesting consequences \cite{jamil1}. 
The validity of 
GSL of thermodynamics has been investigated in the cosmological scenario where dark energy interacts with both dark matter and radiation \cite{jamil2}. 
It has been shown that the GSL 
is always and generally valid, independently of the specific interaction form of the fluids equation of state 
and of the background geometry. 
In addition, viscous dark energy and the GSL of thermodynamics~\cite{Setare:2010zz} and thermodynamics of viscous dark energy in the Rundall-Sundram II~\cite{Randall:1999vf} braneworld~\cite{Setare:2011dt} have been investigated. 
Furthermore, thermodynamic description of the interacting new agegraphic~\cite{Wei:2007ty} dark energy~\cite{Sheykhi:2009ui}, thermodynamic interpretation of the interacting holographic dark energy model~\cite{Li:2004rb, Elizalde:2005ju, Nojiri:2005pu, Amendola:1999er, Zimdahl:2001ar, Chimento:2012fh} in a non-flat universe~\cite{Setare:2008bb}, and the holographic model of dark energy as well as thermodynamics of non-flat accelerated expanding universe~\cite{Setare:2006vz} have been explored. 

The power-law correction to entropy which appears in
dealing with the entanglement of quantum fields in and
out the horizon is given by \cite{power}
\begin{equation}\label{1b}
S_A=\frac{A}{4}\Big(1-K_\alpha A^{1-\frac{\alpha}{2}} \Big)\,,
\end{equation}
where $\alpha$ is a dimensionless constant and a free parameter, and
\begin{equation}\label{1c}
A=4\pi R_h^2, \ \ \ K_\alpha=\frac{\alpha(4\pi)^{\frac{\alpha}{2}-1}}{(4-\alpha)r_c^{2-\alpha}}\,,
\end{equation}
where $r_c$ is a cross-over scale, $R_h$ is the radius, and $A$ is the area of the cosmological horizon. For entropy to be a well-defined quantity, we require $\alpha>0$. The second term in (\ref{1b}) can be regarded as a power-law correction to the area law, resulting from entanglement, when the wavefunction
of the field is chosen to be a superposition of
ground state and exited state \cite{p1}. Several aspects of power-law corrected entropy (\ref{1b}) have been studied in the literature including the GSL \cite{jamil3}, power-law entropy corrected models of dark energy \cite{jamil4} 
(for reviews on dark energy, see, e.g.,~\cite{DE-Reviews}).

The quantum
corrections provided to the entropy-area relationship leads to the curvature correction in the Einstein-Hilbert action
and vice versa.
The logarithmic corrected entropy is \cite{jamil5}
\begin{equation}\label{1a}
S_A=\frac{A}{4}+\beta\log\Big( \frac{A}{4} \Big)+\gamma\,.
\end{equation}
These
corrections arise in the black hole entropy in loop quantum gravity due to thermal equilibrium fluctuations and quantum fluctuations \cite{lqg}.
It has been shown \cite{log} that in a (super) accelerated universe the GSL is valid whenever $\beta(<)>0$ leading to a (negative) positive contribution from logarithmic correction to the entropy. In case of super acceleration the temperature of the dark energy is obtained to be less or equal to the Hawking temperature. Using the corrected entropy-area relation motivated by the loop quantum gravity, 
the validity of the GSL in the FRW universe filled with an interacting viscous dark energy with dark matter and radiation was examined. 
It has been found 
that the GSL is always satisfied throughout the history of the universe for any spatial curvature regardless of the dark energy model.

However in $f(T)$ gravity, the entropy-area relation gets a modification 
like $A\rightarrow A(1+f_T)$, assuming $f_{TT}<<1$ \cite{li}. 
Here $f_T\equiv\frac{df}{dT}$, $f_{TT}\equiv\frac{d^2f}{dT^2}$, 
and $T$ is the torsion scalar. 
The logarithmic corrected entropy becomes
\begin{equation}\label{1d}
S_A=\frac{A(1+f_T)}{4}+\beta\log\Big( \frac{A(1+f_T)}{4} \Big)+\gamma\,.
\end{equation}
For the entropy to be a well-defined quantity, we must have $f_T>0$.
Similarly the power-law corrected entropy becomes
\begin{equation}\label{1e}
S_A=\frac{A(1+f_T)}{4}\Big(1-K_\alpha (A(1+f_T))^{1-\frac{\alpha}{2}} \Big)\,.
\end{equation}
Incidentally, 
thermodynamics with the apparent horizon and the Wald entropy~\cite{Entropy-Wald} in $f(R)$ gravity~\cite{ 
Jacobson:1993vj, Cognola:2005de, 
Bamba:2009ay, Bamba:2010kf}, 
various modified gravitational theories~\cite{Bamba:2009gq, BGT-Thermodynamics}, and $f(T)$ gravity~\cite{Bamba:2011pz} 
has been explored. 

The plan of the paper is as follows: In section II, we write down the Friedmann equations and the laws of thermodynamics for our further use. In section III, we study the GSL 
using the classical Bekenstein-Hawking entropy-area relation with the apparent and event horizons. In sections IV and V, we perform similar analysis for the power-law and logarithmic corrected entropy-area relations. In section VI, we discuss our results. In all sections, we choose units $c=1=G$.

\section{Basic equations}

\subsection{$f(T)$ gravity}
If we accept the equivalence principle, we must work with a curved manifold for the construction of a gauge theory for gravitational field. It is not necessary to use the Riemannian manifolds. The general form of a gauge theory for gravity, with metric, non-metricity and torsion can be constructed easily \cite{smalley}. If we relax the non-metricity, our theory is defined on Weitzenb\"{o}ck spacetime, with torsion and with zero local Riemann tensor. In this theory, which is called teleparallel gravity, we use a non-Riemannian spacetime manifold. The dynamics of the metric determined using the torsion $T$. The basic quantities in teleparallel or the natural extension of it, namely $f(T)$ gravity is the vierbein (tetrad) basis  $e^{i}_{\mu}$ \cite{ff,linder}. This basis is an orthonormal, coordinate free basis, defined by the following equation
\begin{eqnarray}\nonumber
g_{\mu\nu}=e_{\mu}^{i}e_{\nu}^j \eta_{ij}\,.
\end{eqnarray}
This tetrad basis must be orthonormal and $\eta_{ij}$ is the Minkowski flat tensor. It means that $e^{i}_{\mu}e^{\mu }_j=\delta^i_{j}$.
One suitable form of the action for $f(T)$ gravity in Weitzenb\"{o}ck 
spacetime is given by \cite{darabi}
\begin {equation}\label{a-1}
S=\int d^{4}xe\Big(\frac{1}{16\pi}[T+f(T)]+L_{m}\Big)\,,
\end{equation}
where  $f$ is an arbitrary function, $e=\det(e^{i}_{\mu})$. The dynamical quantity of the model is the scalar torsion $T$ and the matter Lagrangian $L_m$.
 Here $T$ is defined by
\begin{equation}\nonumber
T=S^{\:\:\:\mu \nu}_{\rho} T_{\:\:\:\mu \nu}^{\rho}\,,
\end{equation}
with 
$$
T_{\:\:\:\mu \nu}^{\rho}=e_i^{\rho}(\partial_{\mu}
e^i_{\nu}-\partial_{\nu} e^i_{\mu})\,,
$$
$$
S^{\:\:\:\mu \nu}_{\rho}=\frac{1}{2}(K^{\mu
\nu}_{\:\:\:\:\:\rho}+\delta^{\mu}_{\rho} T^{\theta
\nu}_{\:\:\:\theta}-\delta^{\nu}_{\rho} T^{\theta
\mu}_{\:\:\:\theta})\,,
$$
where the asymmetric tensor (which is also called the contorsion tensor) $K^{\mu \nu}_{\:\:\:\:\:\rho}$ reads 
$$
K^{\mu \nu}_{\:\:\:\:\:\rho}=-\frac{1}{2}(T^{\mu
\nu}_{\:\:\:\:\:\rho}-T^{\nu \mu}_{\:\:\:\:\:\rho}-T^{\:\:\:\mu
\nu}_{\rho})\,.
$$
The equation of motion derived from the action, by varying with respect to the $e^{i}_{\mu}$, 
is given by 
\begin{eqnarray}\nonumber
e^{-1}\partial_{\mu}(e S^{\:\:\:\mu
\nu}_{i})(1+f_T)-e_i^{\:\lambda}T_{\:\:\:\mu
\lambda}^{\rho}S^{\:\:\:\nu \mu}_{\rho}f_T\\\nonumber +S^{\:\:\:\mu
\nu}_{i}\partial_{\mu}(T)f_{TT}-\frac{1}{4}e_{\:i}^{\nu}
(1+f(T))=4 \pi G e_i^{\:\rho}T_{\rho}^{\:\:\nu}\,,
\end{eqnarray}
where 
$T_{\mu\nu}$ is the energy-momentum tensor for matter sector of the Lagrangian $L_m$, defined by $$T_{\mu\nu}=-\frac{2}{\sqrt{-g}}\frac{\delta\Big(\int d^4x\sqrt{-g}L_m \Big)}{\delta g^{\mu\nu}}.$$
It is a straightforward calculation to show that this equation of motion is reduced to the Einstein gravity when $f(T)=0$. Indeed, 
this is the equivalence between the teleparallel theory and the Einstein gravity \cite{T}.

We take 
a spatially flat homogeneous and isotropic 
Friedmann-Lema\^{i}tre-Robertson-Walker (FLRW) 
spacetime
\begin{equation}\label{1}
ds^2=-dt^2+a^2(t)[dr^2+r^2(d\theta^2+\sin^2\theta d\phi^2)]\,, 
\end{equation}
where $a(t)$ is the scale factor and 
$d\theta^2+\sin^2\theta d\phi^2 \equiv d \Omega^2$ is the metric of 
two-dimensional sphere with unit radius. 
The Friedmann equations in $f(T)$ gravity are
\begin{eqnarray}
H^2&=&\frac{8\pi}{3}\rho-\frac{1}{6}f-2H^2f_T\,,\label{2}\\
\dot H &=& -\frac{4\pi (\rho+p)}{1+f_T-12H^2f_{TT}}\,.\label{3}
\end{eqnarray}
Here $H=\dot{a}/a$ is the Hubble parameter and the dot denotes the 
time derivative of $\partial/\partial t$. 
In addition, 
$\rho=\sum_i\rho_i $ and $p=\sum_i p_i$ ($i=1\ldots n$) are the total energy density and pressure of $n$ cosmic fluids, respectively. For the FLRW metric (\ref{1}), the trace of the torsion tensor is $T=-6H^2$, and hence 
Eqs.~(\ref{2}) and (\ref{3}) are simplified as
\begin{eqnarray}
H^2&=&\frac{1}{1+2f_T}\Big( \frac{8\pi}{3}\rho-\frac{f}{6}  \Big)\,,\label{4}\\
\dot H &=& -4\pi \Big( \frac{\rho+p}{1+f_T+2Tf_{TT}}  \Big)\,.\label{5}
\end{eqnarray}
We will work under the assumption $f_{TT}<<1$. Thus 
Eq. (\ref{5}) is reduced 
to
\begin{eqnarray}
\dot H &\simeq& -4\pi \Big( \frac{\rho+p}{1+f_T}  \Big)\,.\label{5a}
\end{eqnarray}
Moreover, 
the energy conservation equation is given by
\begin{equation} \label{5a}
\dot\rho+3H(\rho+p)=0\,.
\end{equation}

\subsection{Generalized second law (GSL)} 

The GSL 
of thermodynamics for black holes states that entropy of a black hole added with the entropy of the background universe must be non-decreasing. In other words, the generalized entropy function must be positive definite: $S\equiv\sum_i S_i+S_h\geq0$, where $S_i$ and $S_h$ are the entropies of individual component and the event horizon. It is proved in Ref.~\cite{li} that the usual first law of thermodynamics does not hold in $f(T)$ gravity and an extra term due to `entropy production' $S_p$ is introduced to the first law. Therefore to study the GSL, we use the `modified first law of thermodynamics' \cite{li}
\begin{equation}
T_idS_i=d(\rho_iV)+p_idV-T_idS_p\,, \label{6}
\end{equation}
where $T_i$ is the temperature and $S_i$ is the entropy of the $i$th component of the fluid, and also $V=\frac{4\pi}{3}R_h^3$ is the volume of the horizon.

We assume the $n$-component fluid to be interacting and exchange energy. 
Although this assumption is purely phenomenological, 
it helps in resolving certain problems in cosmology including 
a coincidence problem \cite{rahaman} and is also consistent with the astrophysical observations \cite{obs}. The continuity equation (\ref{5a}) for each fluid reads
\begin{equation} \label{7}
\dot\rho_i+3H(\rho_i+p_i)=Q_i\,,
\end{equation}
where $Q_i$ are the interaction terms, collectively satisfying $\sum_i Q_i=0$. 
{}From Eqs. (\ref{6}) and (\ref{7}), we get
\begin{equation}\label{8}
\dot S_i=\frac{4\pi}{3}R_h^3\frac{Q_i}{T_i}+4\pi R_h^2(\dot R_h-HR_h)(\frac{\rho_i+p_i}{T_i})-\dot S_p\,.
\end{equation} 
We suppose 
the thermal equilibrium, i.e., $T_i=T_H$, and thus
\begin{equation}\label{8a}
\dot S_I+\dot S_p=-\frac{R_h^2}{T_H}(\dot R_h-HR_h)\dot H(1+f_T)\,,
\end{equation}
where $S_I=\sum_i S_i$ is the total internal entropy of $n-$ fluids. The form of GSL in this work 
reads $$ \dot S_{tot}=\dot S_I+\dot S_p+\dot S_h\geq0\,. $$ Here we introduce two definitions of cosmological horizons from the literature: the \textit{dynamical apparent horizon} is a null surface with vanishing expansion. In the flat FLRW universe, it is $R_A=H^{-1}$ (which is also called Hubble horizon) \cite{cai}. 
The second interesting case is the \textit{future event horizon} which is the distance that light travels from the present time to 
infinity and is defined as
\begin{equation}
R_E=a(t)\int\limits_t^\infty\frac{dt'}{a(t')}\,,\ \   \dot{R}_E=HR_E-1\,.
\end{equation} 

\section{GSL with Bekenstein-Hawking entropy-area relation}

We consider the form of entropy in $f(T)$ gravity
\begin{equation}\label{9} 
S_A=S_A( X)\,, \ \ X\equiv\frac{A(1+f_T)}{4}\,.
\end{equation}
Differentiating (\ref{9}) with respect to (w.r.t.) $t$, we get
\begin{equation}\label{10}
\dot S_A=\Big[ \frac{(1+f_T)\dot A}{4}+\frac{A}{4}f_{TT}\dot T  \Big]\frac{dS}{dX}\,. 
\end{equation}
Ignoring $f_{TT}$, we obtain
\begin{equation}\label{11}
\dot S_A\simeq\frac{(1+f_T)\dot A}{4}\frac{dS}{dX}|_{X}\,.
\end{equation}

Bekenstein-Hawking entropy in $f(T)$ gravity is described by 
\begin{equation}\label{12}
S_A(X)=X\,.
\end{equation}
Hence (\ref{11}) implies
\begin{equation}\label{13}
\dot S_A=\frac{(1+f_T)\dot A}{4}\,.
\end{equation}

\subsection{
Use of the apparent horizon}

It has been shown \cite{zhou} that in an accelerating universe, the GSL holds only in the case that 
the boundary surface is the apparent horizon, 
not in the case of the event horizon. 
This suggests that event horizon is not a physical boundary from the point of view of thermodynamics.

We examine 
the dynamical apparent horizon \cite{cai}
\begin{equation}\label{14}
R_A=\frac{1}{H}\,.
\end{equation}
The apparent horizon $R_A$ is a marginally trapped surface with vanishing expansion and is determined from the condition $g^{ij}\partial_i\tilde{r}\partial_j\tilde{r}=0$, where $\tilde{r}=r(t)a(t)$ and $i,j=0,1$ \cite{cao}. Assuming $A=4\pi R_A^2$, we have
\begin{equation}\label{15}
\dot S_A=-2\pi\frac{\dot H}{H^3}(1+f_T)\,.
\end{equation}
Clearly by substituting $f=0,$ we obtain results for the usual Hawking entropy.
The form of the GSL expression here becomes
\begin{equation}\label{15a}
\dot S_{tot}\equiv\dot S_A+\dot S_I+\dot S_p=2\pi\frac{\dot H^2}{H^5}(1+f_T)\geq0\,,
\end{equation}
where we have assumed the thermal equilibrium for our dynamical system $T_i=T_H$ and used 
$T_H =\frac{H}{2\pi}$ \cite{aku}. 

We notice that the assumption of the thermal equilibrium in cosmological setting is very ideal, because 
major components of the universe including dark matter, dark energy and radiation (cosmic microwave background (CMB) and neutrinos) have entirely different temperatures \cite{lima}. 
However it has been found that the contribution of the heat flow between dark energy and dark matter for the GSL in the nonequilibrium 
thermodynamics is very small as $O(10^{-7})$ \cite{kk}. Therefore the equilibrium thermodynamics
is still preserved. Further, if there is any thermal difference between the fluids and the horizon, the transfer of energy across the horizon might change the geometry of the horizon and the FLRW spacetime \cite{log}. 
We mention that 
thermodynamics with the apparent horizon and the Wald entropy 
in a model of $f(R)$ gravity realizing the crossing of the phantom divide~\cite{Bamba:2008hq}, which can occur also in $f(T)$ gravity~\cite{BGLL-BGL}, 
has been studied~\cite{Bamba:2009ay}. 
In addition, various cosmological aspects in $f(T)$ gravity 
have widely been examined in, e.g., Refs.~\cite{Jamil:2012fs, Jamil:2012ju, 
Jamil:2012vb, Jamil:2012nm} 
and those in teleparallelism in, for example, Refs.~\cite{Wei:2011yr, Xu:2012jf, Gu:2012ww, Bamba:2012mi}.

\subsection{
Use of the event horizon}

Time derivative of the entropy of the future event horizon ($S_A=\pi R_E^2(1+f_T)$) is given by
\begin{equation}\label{16}
\dot S_A=2\pi R_E\dot R_E(1+f_T)\,.
\end{equation}
The total entropy for the GSL becomes 
\begin{equation}\label{15a}
\dot S_{tot}=2\pi R_E\dot R_E (1+f_T)+\frac{2\pi R_E^2}{bH}\dot H (1+f_T)\geq0\,,
\end{equation}
where we have provided 
the thermal equilibrium for our dynamical system
and used $T_H=\frac{bH}{2\pi}$ with 
$b$ being a constant \cite{aku}.
Using the relation $\dot{R_{E}}=HR_E-1$, we rewrite (\ref{15a}) in
the following form
\begin{equation}\label{log1}
\dot S_{tot}\equiv2\pi R_E^2(1+f_T)\frac{d}{dt}\Big[\log(R_E
H^{\frac{1}{b}})\Big]\geq0\,.
\end{equation}
Since $f_T>0$, there is only needed to check the positivity of the
logarithmic term. We choose the following set of the parameters \cite{odintsov}: 
\begin{eqnarray} 
a(t)=a_0 (t_s-t)^{-n}, \ H=\frac{n}{t_s-t}, \
 R_E=\frac{t_s-t}{n+1},
\end{eqnarray}
where $a_0>0$, $n>0$ and $t_s>t>0$ with $t_s$ being the time of occurrence of a future cosmic singularity like big rip. 
By the direct substitution of these equations into (\ref{log1}), we find that the following range of $b$ must be valid 
\begin{eqnarray}
b\leq1.
\end{eqnarray}

\section{GSL of thermodynamics with power-law entropy correction}
The form of entropy with the power-law correction term is given by
\begin{equation}\label{16a}
S(X)=X[1-K_\alpha(4X)^{1-\frac{\alpha}{2}}]\,,
\end{equation}
whose time derivative becomes
\begin{eqnarray}\label{16b}
\dot S&\simeq& \frac{\dot A(1+f_T)}{4}\Big[ 1-K_\alpha \Big(A(1+f_T)\Big)^{1-\frac{\alpha}{2}}\nonumber\\&&-K_\alpha\Big(  1-\frac{\alpha}{2}\Big)\Big(A(1+f_T)\Big)^{1-\frac{\alpha}{2}} \Big]\,.
\end{eqnarray}

\subsection{Using the apparent horizon}

Taking time derivative of the entropy with the power-law correction and using the apparent horizon, we get  
\begin{eqnarray}\label{16c}
&&
\dot S_A=-2\pi \frac{\dot H}{H^3}(1+f_T)\Big[  1- 
\nonumber \\ 
&&
\hspace{10mm}
K_\alpha\Big(2-\frac{\alpha}{2}\Big)
\Big( \frac{4\pi}{H^2}(1+f_T) \Big)^{1-\frac{\alpha}{2}} \Big]\,.
\end{eqnarray}
Moreover the total entropy for the GSL is obtained by adding (\ref{16c}) to 
(16), we acquire
\begin{eqnarray}\label{16c1}
&&
\hspace{-5mm}
\dot S_{tot}=2\pi \frac{\dot H}{H^3}(1+f_T)\Big[ \frac{\dot H}{H^2}
\nonumber \\ 
&&
\hspace{5mm}
 {}+K_\alpha\Big(2-\frac{\alpha}{2}\Big)\Big( \frac{4\pi}{H^2}(1+f_T) \Big)^{1-\frac{\alpha}{2}} \Big]\geq0\,.
\end{eqnarray}
We rewrite (\ref{16c1}) in the following form
\begin{eqnarray}
&&
\dot S_{tot}=2\pi
\frac{\dot{H}}{H^3}(1+f_T)\Big[\frac{\dot{H}}{H^2}
\nonumber \\ 
&&
\hspace{10mm}
{}+\frac{\alpha}{2}(Hr_c)^{\alpha-2}
(1+f_T)^{1-\frac{\alpha}{2}}\Big]\geq0\,.
\end{eqnarray}
There exist two special cases:
In Quintessence [the non-Phantom] ($\dot{H}<0$) phase, 
\begin{itemize}
\item 
for $\alpha>0$, we find 
$\dot{H}\leq-\frac{\alpha
H^2}{2}(Hr_c)^{\alpha-2}(1+f_T)^{1-\frac{\alpha}{2}}$\,.
\item 
for $\alpha<0$, 
we obtain 
$\dot S_{tot}\geq0$ for any 
$H, \dot{H}, \alpha$.
\end{itemize}
On the other hand, in the Phantom phase ($\dot{H}>0$), 
\begin{itemize}
\item 
for $\alpha<0$, we acquire 
$\dot{H}\geq-\frac{\alpha
H^2}{2}(Hr_c)^{\alpha-2}(1+f_T)^{1-\frac{\alpha}{2}}$\,.
\item 
for 
$\alpha>0$, we have 
$\dot S_{tot}\geq0$ for any 
$H, \dot{H}, \alpha$\,.
\end{itemize}

\subsection{Using the event horizon}

For the power-law entropy, the time derivative of the horizon entropy is 
written as 
\begin{eqnarray}\label{16d}
&&
\dot S_E=2\pi R_E\dot R_E(1+f_T)\Big[ 1 
\nonumber \\ 
&&
\hspace{10mm}
-K_\alpha (2-\frac{\alpha}{2})\Big(4\pi  R_E^2(1+f_T) \Big)^{1-\frac{\alpha}{2}}\Big]\,,
\end{eqnarray}
and the form of the GSL becomes
\begin{eqnarray}\label{16d1}
&&
\hspace{-10mm}
\dot S_{tot}=2\pi R_E\dot R_E(1+f_T)\Big[ 1-K_\alpha\Big (2-\frac{\alpha}{2}\Big) \nonumber \\
&& 
\hspace{-10mm}
{}\times \Big(4\pi  R_E^2(1+f_T) \Big)^{1-\frac{\alpha}{2}}\Big]
+\frac{2\pi R_E^2}{bH}\dot H(1+f_T)\geq0\,.
\end{eqnarray}
We rewrite (\ref{16d1}) in the following form
\begin{eqnarray}
&&
\dot S_{tot}=2\pi R_E (1+f_T)\Big[\dot{R_E}(1-\frac{\alpha}{2}(R_E
r_c)^{2-\alpha}) 
\nonumber\\
&&
\hspace{10mm}
{}\times 
(1+f_T)^{1-\frac{\alpha}{2}}
+\frac{\dot{H}}{bH}R_E\Big]\geq0.
\end{eqnarray}
\begin{itemize}
\item 
For the Quintessence phase: $\dot{H}<0$, $\dot{R_E}>0$, we have
\begin{eqnarray}
\frac{\dot{R_E}}{R_E}\geq-\frac{\dot{H}}{bH}\Big(1-\frac{\alpha}{2}(R_E
r_c)^{2-\alpha}(1+f_T)^{1-\frac{\alpha}{2}}\Big)\,. 
\end{eqnarray}
\item 
For the Phantom phase: $\dot{H}>0$, $\dot{R_E}<0$, we acquire 
\begin{eqnarray}
\frac{\dot{R_E}}{R_E}\leq-\frac{\dot{H}}{bH}\Big(1-\frac{\alpha}{2}(R_E
r_c)^{2-\alpha}(1+f_T)^{1-\frac{\alpha}{2}}\Big)\,.
\end{eqnarray}
\end{itemize}

\section{GSL of thermodynamics with logarithmic correction}

The entropy with the logarithmic correction is expressed as 
\begin{equation}\label{17}
S(X)=X+\beta\log (X)+\gamma\,,
\end{equation}
where $\beta$ and $\gamma$ are constants. Upon the differentiation w.r.t $t$, 
we get
\begin{equation}\label{18}
\dot S\simeq \frac{\dot A(1+f_T)}{4}\Big( 1+\frac{4\beta H^2}{A(1+f_T)} \Big)\,.
\end{equation}

\subsection{
Case of the apparent horizon}
For the apparent horizon, (\ref{18}) reduces to
\begin{equation}\label{19}
\dot S_A= \frac{-2\pi\dot H}{H^3}(1+f_T)\Big( 1+\frac{4\beta H^2}{\pi(1+f_T)} \Big)\,,
\end{equation}
while the total entropy for the GSL reads
\begin{equation}\label{19a}
\dot S_{tot}= \frac{-2\pi\dot H}{H^3}(1+f_T)\Big( \frac{\beta H^2}{\pi(1+f_T)}-\frac{\dot H}{H^2} \Big)\geq0\,.
\end{equation}
Let us concentrate on 
only two special cases of interest:
In the Quintessence phase ($\dot{H}<0$), 
\begin{itemize}
\item 
for $\beta>0$, we find 
$\dot S_{tot}\geq0$ in any time.
\item 
for $\beta<0$, we obtain 
$\dot{H}\leq\frac{\beta H^4}{\pi (1+f_T)}$ in any time.
\end{itemize}
In the Phantom phase ($\dot{H}>0$): 
\begin{itemize}
\item 
for $\beta<0$, we have 
$\dot S_{tot}\geq0$ in any time.
\item 
for $\beta>0$, we acquire 
$\dot{H}\geq\frac{\beta H^4}{\pi (1+f_T)}$ in any time.
\end{itemize}

\subsection{
Case of the event horizon}
The definition of entropy for the comic event horizon with the logarithmic correction converts (\ref{18}) to
\begin{equation}\label{20}
\dot S_A= 2\pi R_E\dot R_E(1+f_T)\Big( 1+\frac{\beta}{\pi R_E^2(1+f_T)} \Big)\,,
\end{equation}
and in this case the corresponding form of the GSL becomes 
\begin{eqnarray}\label{21}
&&
\dot S_{tot} = 2\pi R_E\dot R_E(1+f_T)\Big( 1+\frac{\beta}{\pi R_E^2(1+f_T)} \Big)\nonumber\\
&&
\hspace{10mm}
{}+\frac{2\pi R_E^2}{bH}\dot H (1+f_T)\geq0.
\end{eqnarray}
\begin{itemize}
\item 
In the Quintessence era, $\dot{H}<0$, $\dot{R_E}>0$, we find
\begin{eqnarray}
\frac{\dot{R_E}}{R_E}\geq-\frac{\dot{H}}{bH}\Big(1+\frac{\beta}{\pi
R_E^2(1+f_T)}\Big)^{-1}\,. 
\end{eqnarray}
\item 
In the Phantom era, $\dot{H}>0$, $\dot{R_E}<0$, we obtain
\begin{eqnarray}
\frac{\dot{R_E}}{R_E}\leq-\frac{\dot{H}}{bH}\Big(1+\frac{\beta}{\pi
R_E^2(1+f_T)}\Big)^{-1}\,. 
\end{eqnarray}
\end{itemize}

\section{Discussion}
In the present paper, we have investigated the validity of the GSL in $f(T)$ gravity. We have used the power-law and logarithmic corrected forms of entropy for the cosmological horizon. 
An important point we have made is that 
the classical entropy-area law gets modification 
after adding the term $(1+f_T)$ to its expression. Since $f_T>0$, it implies that the usual entropy of the horizon increases by the corresponding factor $(1+f_T)$, although 
the result is analogous to that in $f(R)$ gravity. 
However the crucial difference between thermodynamics in $f(R)$ and $f(T)$ gravities is that the first law of thermodynamics does not hold in $f(T)$ unlike $f(R)$. 
Nevertheless, 
after adding an entropy production term, the modified form of the first law 
becomes valid for thermodynamic 
studies. 
We have performed our analysis for the apparent and event horizons separately. 
In the usual Bekenstein-Hawking area law in $f(T)$ cosmology, 
for both of the Quintessence and Phantom regimes, we have found general conditions for the validity of the GSL. 
In the case of the event horizon, we have explicitly demonstrated that if the constant of the proportionality is $b\leq1$, then the GSL remains valid. 
For the power law corrected form of the entropy, again we have shown that 
for either the Phantom or Quintessence dominated eras, the GSL is valid for any value of $\alpha$. 
In the case of the event horizon, we have found that the validity of the GSL depends on the rate of the change $\frac{d\log(R_E)}{dt}$. 
For the logarithmic corrected form of the entropy, we have found that with the apparent horizon in both of the Phantom and Quintessence regimes, the GSL is valid. 
In the case of the event horizon, same as the power law corrected case, the validity of the GSL depends on the time derivative of the $\log(R_E)$. 
Our results are generally valid regardless the form of $f(T)$. 
As a result, our work has clarified some thermodynamic 
features of the $f(T)$ gravity.

Finally, we remark that 
the modification of the FLRW equations in the presence of the entropy 
corrections is an interesting and open problem, 
similarly to that, for example, in entropic cosmology or warped codimension-two braneworld~\cite{Chen:2008sn}. 
Indeed, which kind of the modifications are possible in $f(T)$ gravity 
has not been understood yet. 
Thus, it would be meaningful to indicate that 
there can exist possible extensions of the FLRW equations 
under the entropy corrections in $f(T)$ gravity.



\begin{thebibliography}{99}

\bibitem{barton} J. Ellis, N. E. Mavromatos, D.V. Nanopoulos, Phys. Lett. B {\bf 276}, 56 (1992); S. Mignemi, N.R. Stewart,  Phys. Rev. D {\bf 47}, 5259 (1993); B. Zwiebach, A First Course on String Theory, (Cambridge Universe Press, 2004); C. Rovelli, Quantum Gravity  (Cambridge Universe Press).

\bibitem{hawking} S. W. Hawking, Commun. Math. Phys. {\bf 43}, 199 (1975); J. D. Bekenstein, Phys. Rev. D {\bf 7}, 2333 (1973); J. M. Bardeen, B. Carter and S. W. Hawking, Commun. Math. Phys. {\bf 31}, 161 (1973).

\bibitem{paddy} T. Padmanabhan, Phys. Rept. {\bf 406}, 49 (2005); Rep. Prog. Phys. {\bf 73}, 046901 (2010).

\bibitem{Jacobson:1995ab}
  T.~Jacobson,
  Phys.\ Rev.\ Lett.\  {\bf 75}, 1260 (1995).

\bibitem{Eling:2006aw}
  C.~Eling, R.~Guedens and T.~Jacobson,
  Phys.\ Rev.\ Lett.\  {\bf 96}, 121301 (2006).

\bibitem{Elizalde:2008pv}
  E.~Elizalde and P.~J.~Silva,
  Phys.\ Rev.\  D {\bf 78}, 061501 (2008).

\bibitem{Brustein:2009hy}
  R.~Brustein and M.~Hadad,
  Phys.\ Rev.\ Lett.\  {\bf 103}, 101301 (2009).

\bibitem{cai} R-G Cai, S. P. Kim, JHEP {\bf 0502}, 050 (2005).

\bibitem{akbar1} M. Akbar, R.-G. Cai, Phys. Lett. B {\bf 635}, 7 (2006); A. Sheykhi,  Eur. Phys. J. C {\bf 69}, 265 (2010); A. Sheykhi, B. Wang, R-G. Cai, Phys. Rev. D {\bf 76}, 023515 (2007).


\bibitem{Reviews-F(R)}
%
  S.~Nojiri and S.~D.~Odintsov,
  Phys.\ Rept.\  {\bf 505}, 59 (2011);\ 
%
 eConf {\bf C0602061}, 06 (2006)
 [Int.\ J.\ Geom.\ Meth.\ Mod.\ Phys.\  {\bf 4}, 115 (2007)]
 [arXiv:hep-th/0601213];\ 
%
 T.~P.~Sotiriou and V.~Faraoni,
 Rev.\ Mod.\ Phys.\  {\bf 82}, 451 (2010);\ 
%
  A.~De Felice and S.~Tsujikawa,
  Living Rev.\ Rel.\  {\bf 13}, 3 (2010);\ 
%
S.~Capozziello and V.~Faraoni,
\textit{Beyond Einstein Gravity}
(Springer, 2010);\ 
%
  S.~Tsujikawa,
  Lect.\ Notes Phys.\  {\bf 800}, 99 (2010)
  [arXiv:1101.0191 [gr-qc]];\ 
%
  T.~Clifton, P.~G.~Ferreira, A.~Padilla and C.~Skordis,
  Phys.\ Rept.\  {\bf 513}, 1 (2012);\ 
%
  S.~Capozziello and M.~De Laurentis,
  Phys.\ Rept.\  {\bf 509}, 167 (2011);\ 
%
  T.~Harko and F.~S.~N.~Lobo,
  arXiv:1205.3284 [gr-qc];\ 
%
  S.~Capozziello, M.~De Laurentis and S.~D.~Odintsov,
  Eur.\ Phys.\ J.\ C {\bf 72}, 2068 (2012). 
%

\bibitem{page} V. P. Frolov, D. N. Page, Phys. Rev. Lett. {\bf 71}, 3902 (1993); S. Mukohyama, Phys. Rev. D {\bf 56}, 2192 (1997); A.C. Wall, JHEP {\bf 0906}, 021 (2009).

\bibitem{shim} T. Shimomura, T. Okamura, T. Mishima, H. Ishihara, Phys. Rev. D {\bf 62}, 044036 (2000); Addendum-ibid. D {\bf 65}, 107502 (2002).

\bibitem{wald} S. Gao, R. M. Wald, Phys. Rev. D {\bf 64}, 084020 (2001).

\bibitem{quantum} A. Hosoya, A. Carlini, T. Shimomura,  Phys. Rev. D {\bf 63}, 104008 (2001); D. D. Song, E. Winstanley, Int. J. Theor. Phys. {\bf 47}, 1692 (2008);
 Q-R. Zhang, Int. J. Mod. Phys. E {\bf 17}, 531 (2008).

\bibitem{shells} S. He, H. Zhang, JHEP {\bf 0712}, 052 (2007).

\bibitem{violate} L.H. Ford, T.A. Roman, Phys. Rev. D {\bf 64}, 024023 (2001).

\bibitem{string} R. Brustein, S. Foffa, R. Sturani, Phys. Lett. B {\bf 471}, 
352 (2000); R. Brustein, Phys. Rev. Lett. {\bf 84}, 2072 (2000); 
B. Wang, E. Abdalla, Phys. Lett. B {\bf 471}, 346 (2000).

\bibitem{rip} J.A. de Freitas Pacheco, J.E. Horvath, Class. Quant. Grav. {\bf 24}, 5427 (2007); K. Nouicer, Int. J. Mod. Phys. D {\bf 20}, 233 (2011).

\bibitem{hl} M. Jamil, E. N. Saridakis, M. R. Setare, JCAP {\bf 1011}, 032 (2010).

\bibitem{sad} H. M. Sadjadi, Europhys. Lett. {\bf 92}, 50014 (2010); 
Phys. Rev. D {\bf 76}, 104024 (2007); S-F. Wu, B. Wang, G-H. Yang, P.-M. Zhang,
Class. Quant. Grav. {\bf 25}, 235018 (2008); A. Sheykhi, B. Wang, Phys. Lett. B {\bf 678}, 434 (2009); 
N. Mazumder, S. Chakraborty, Astrophys. Space Sci. {\bf 332}, 509 (2011).

\bibitem{f} H. Farajollahi, A. Salehi, F. Tayebi, Can. J. Phys. {\bf 89}, 915 (2011).

\bibitem{davies} P. C. W. Davies, T. M. Davis, 
Found. Phys. {\bf 32}, 1877 (2002); 
G. Izquierdo, D. Pavon, Phys. Lett. B {\bf 633}, 420 (2006).

\bibitem{sad1} H. Mohseni Sadjadi, Phys. Rev. D {\bf 73}, 063525 (2006).

\bibitem{shit} G. Izquierdo, D. Pavon, Phys. Lett. B {\bf 639}, 1 (2006).

\bibitem{sad2} H.M. Sadjadi, Phys. Lett. B {\bf 645}, 108 (2007).

\bibitem{setare} M. R. Setare, Phys. Lett. B {\bf 641}, 130 (2006).

\bibitem{setare1} M. R. Setare, JCAP {\bf 0701}, 023 (2007).

\bibitem{jamil1} M. Jamil, M. Akbar, Gen. Rel. Grav. {\bf 43}, 1061 (2011).

\bibitem{jamil2} M. Jamil, E.N. Saridakis, M.R. Setare, Phys. Rev. D {\bf 81}, 023007 (2010).

\bibitem{Setare:2010zz} 
  M.~R.~Setare and A.~Sheykhi,
  Int.\ J.\ Mod.\ Phys.\ D {\bf 19}, 1205 (2010).

\bibitem{Randall:1999vf} 
  L.~Randall and R.~Sundrum,
  Phys.\ Rev.\ Lett.\  {\bf 83}, 4690 (1999). 

\bibitem{Setare:2011dt} 
  M.~R.~Setare and A.~Sheykhi,
  Int.\ J.\ Mod.\ Phys.\ D {\bf 19}, 171 (2010). 

\bibitem{Wei:2007ty} 
  H.~Wei and R.~-G.~Cai,
  Phys.\ Lett.\ B {\bf 660}, 113 (2008). 

\bibitem{Sheykhi:2009ui} 
  A.~Sheykhi and M.~R.~Setare,
  Mod.\ Phys.\ Lett.\ A {\bf 26}, 1897 (2011). 

\bibitem{Li:2004rb}
  M.~Li,
  Phys.\ Lett.\  B {\bf 603}, 1 (2004).

\bibitem{Elizalde:2005ju}
  E.~Elizalde, S.~Nojiri, S.~D.~Odintsov and P.~Wang,
  Phys.\ Rev.\  D {\bf 71}, 103504 (2005).

\bibitem{Nojiri:2005pu}
  S.~Nojiri and S.~D.~Odintsov,
  Gen.\ Rel.\ Grav.\  {\bf 38}, 1285 (2006).

%
\bibitem{Amendola:1999er}
  L.~Amendola,
  Phys.\ Rev.\  D {\bf 62}, 043511 (2000). 
%

\bibitem{Zimdahl:2001ar}
  W.~Zimdahl, D.~Pavon and L.~P.~Chimento
  Phys.\ Lett.\  B {\bf 521}, 133 (2001). 
%

%
\bibitem{Chimento:2012fh} 
  L.~P.~Chimento and M.~G.~Richarte,
  arXiv:1207.1492 [astro-ph.CO].
%

\bibitem{Setare:2008bb} 
  M.~R.~Setare and E.~C.~Vagenas,
  Phys.\ Lett.\ B {\bf 666}, 111 (2008). 

\bibitem{Setare:2006vz} 
  M.~R.~Setare and S.~Shafei,
  JCAP {\bf 0609}, 011 (2006).

\bibitem{power} A. Sheykhi, M. Jamil, Gen. Rel. Grav. {\bf 43}, 2661 (2011).

\bibitem{p1} S. Das, S. Shankaranarayanan and S. Sur, Phys. Rev. D 
{\bf 77}, 064013 (2008); S. Das, S. Shankaranarayanan and S. Sur,
arXiv:1002.1129; S. Das, S. Shankaranarayanan and S. Sur,
arXiv:0806.0402; N. Radicella, D. Pavon, Phys. Lett. B {\bf 691}, 121 (2010).

\bibitem{jamil3} U. Debnath, S. Chattopadhyay, I. Hussain, M. Jamil, R. Myrzakulov, Eur. Phys. J. C {\bf 72}, 1875 (2012).

\bibitem{jamil4} 
M.U. Farooq, M. Jamil, Canadian J. Phys. {\bf 89}, 1251 (2011); 
E. Ebrahimi, A. Sheykhi, Phys. Scripta {\bf 04}, 045016 (2011); K. Karami, A. Abdolmaleki, N. Sahraei, S. Ghaffari, JHEP {\bf 1108}, 150 (2011).

\bibitem{DE-Reviews}
%
 E.~J.~Copeland, M.~Sami and S.~Tsujikawa,
 Int.\ J.\ Mod.\ Phys.\  D {\bf 15}, 1753 (2006);\ 
%
  R.~R.~Caldwell and M.~Kamionkowski,
  Ann.\ Rev.\ Nucl.\ Part.\ Sci.\  {\bf 59}, 397 (2009);\ 
%
  L.~Amendola and S.~Tsujikawa, 
\textit{Dark Energy}
(Cambridge University press, 2010);\ 
%
  S.~Tsujikawa,
  arXiv:1004.1493 [astro-ph.CO];\ 
%
  M.~Li, X.~D.~Li, S.~Wang and Y.~Wang,
  Commun.\ Theor.\ Phys.\  {\bf 56}, 525 (2011);\ 
%
  M.~Kunz,
  arXiv:1204.5482 [astro-ph.CO];\ 
%
  K.~Bamba, S.~Capozziello, S.~Nojiri and S.~D.~Odintsov,
  Astrophys.\ Space Sci.\  {\bf 342}, 155 (2012).
%

\bibitem{jamil5} R. Banerjee, S. K. Modak, JHEP {\bf 0911}, 073 (2009); S. Banerjee, R. K. Gupta, A. Sen, JHEP {\bf 1103}, 147 (2011); H. Wei, Commun. Theor. Phys. {\bf 52} (2009) 743; H. M. Sadjadi, M. Jamil, Gen. Rel. Grav. {\bf 43}, 1759 (2011); M. Jamil, M.U. Farooq, JCAP {\bf 1003}, 001 (2010); 
M. Akbar, K. Saifullah, Gen. Relativ. Gravit. {\bf 43}, 933 (2011); 
Eur. Phys. J. C {\bf 67}, 205 (2010). 

\bibitem{lqg} C. Rovelli, Phys. Rev. Lett. {\bf 77}, 3288 (1996); A. Ashtekar, J. Baez, A. Corichi, and K. Krasnov, Phys. Rev. Lett. {\bf 80}, 904 (1998); 
A. Ghosh and P. Mitra, Phys. Rev. D {\bf 71}, 027502 (2005); K.A. Meissner, Class. Quant. Grav. {\bf 21}, 5245 (2004); 
A.J.M. Medved and E.C. Vagenas, Phys. Rev. D {\bf 70}, 124021 (2004).

\bibitem{log} H. M. Sadjadi, M. Jamil, Europhys. Lett. {\bf 92}, 69001 (2010).

%
%

\bibitem{li} R.-X. Miao, M. Li, Y-G. Miao, JCAP {\bf 1111}, 033 (2011).

\bibitem{Entropy-Wald}
%
  R.~M.~Wald,
  Phys.\ Rev.\  D {\bf 48}, 3427 (1993);\ 
%
  V.~Iyer and R.~M.~Wald,
  Phys.\ Rev.\  D {\bf 50}, 846 (1994).
%

\bibitem{Jacobson:1993vj}
  T.~Jacobson, G.~Kang and R.~C.~Myers,
  Phys.\ Rev.\  D {\bf 49}, 6587 (1994).

\bibitem{Cognola:2005de}
  G.~Cognola, E.~Elizalde, S.~Nojiri, S.~D.~Odintsov and S.~Zerbini,
  JCAP {\bf 0502}, 010 (2005). 

\bibitem{Bamba:2009ay}
  K.~Bamba and C.~Q.~Geng,
  Phys.\ Lett.\  B {\bf 679}, 282 (2009).

\bibitem{Bamba:2010kf}
  K.~Bamba and C.~Q.~Geng,
  JCAP {\bf 1006}, 014 (2010).

\bibitem{Bamba:2009gq}
  K.~Bamba, C.~Q.~Geng, S.~Nojiri and S.~D.~Odintsov,
  Europhys.\ Lett.\  {\bf 89}, 50003 (2010).

\bibitem{BGT-Thermodynamics}
%
  K.~Bamba, C.~Q.~Geng and S.~Tsujikawa,
  Phys.\ Lett.\  B {\bf 688}, 101 (2010);\ 
%
  Int.\ J.\ Mod.\ Phys.\  D {\bf 20}, 1363 (2011). 
%

\bibitem{Bamba:2011pz}
  K.~Bamba and C.~Q.~Geng,
  JCAP {\bf 1111}, 008 (2011). 

\bibitem{smalley}L. Smalley, Phys. Lett. A {\bf 61}, 436 (1977).

\bibitem{linder} E. V. Linder, Phys. Rev. D {\bf 81}, 127301 (2010) [Erratum-ibid.\  D {\bf 82}, 109902 (2010)].

\bibitem{ff} 	R. Ferraro, F. Fiorini, Phys. Rev. D {\bf 75}, 084031 (2007);
R. Ferraro, F. Fiorini, Phys. Rev. D {\bf 78}, 124019 (2008).

\bibitem{darabi}  
M. R. Setare, F. Darabi, arXiv:1110.3962; 
M. Jamil, D. Momeni, R. Myrzakulov, Eur. Phys. J. C 72, 1959 (2012); 
R. Myrzakulov, Eur. Phys. J. C {\bf 71}, 1752 (2011);
K.K. Yerzhanov, Sh.R. Myrzakul, I.I. Kulnazarov, R. Myrzakulov, 
arXiv:1006.3879; 
%
  P.~Y.~Tsyba, I.~I.~Kulnazarov, K.~K.~Yerzhanov and R.~Myrzakulov,
  Int.\ J.\ Theor.\ Phys.\  {\bf 50}, 1876 (2011);\ 
%
  K.~Bamba, R.~Myrzakulov, S.~Nojiri and S.~D.~Odintsov,
  Phys.\ Rev.\ D {\bf 85}, 104036 (2012);\ 
%
%
  S.~H.~Chen, J.~B.~Dent, S.~Dutta and E.~N.~Saridakis,
  Phys.\ Rev.\  D {\bf 83}, 023508 (2011);\ 
%
  J.~B.~Dent, S.~Dutta and E.~N.~Saridakis,
  JCAP {\bf 1101}, 009 (2011);\ 
%
  Y.~F.~Cai, S.~H.~Chen, J.~B.~Dent, S.~Dutta and E.~N.~Saridakis,
  Class.\ Quant.\ Grav.\  {\bf 28}, 2150011 (2011);\ 
%
  C.~Q.~Geng, C.~C.~Lee, E.~N.~Saridakis and Y.~P.~Wu,
  Phys.\ Lett.\  B {\bf 704}, 384 (2011);\ 
%
  C.~Q.~Geng, C.~C.~Lee and E.~N.~Saridakis,
  JCAP {\bf 1201}, 002 (2012);\ 
%
  P.~A.~Gonzalez, E.~N.~Saridakis and Y.~Vasquez,
  JHEP {\bf 1207}, 053 (2012).

\bibitem{T}
K. Hayashi, T. Shirafuji, Phys. Rev. D {\bf 19}, 3524
(1979); K. Hayashi, T. Shirafuji, Phys. Rev. D {\bf 24},
3312 (1981).

\bibitem{rahaman} M. Jamil, F. Rahaman, Eur. Phys. J. C {\bf 64}, 97 (2009); M. Szydlowski, Phys. Lett. B {\bf 632}, 1 (2006); D. Pavon, W. Zimdahl, Phys. Lett. B {\bf 628}, 206 (2005); M. S. Berger, H. Shojaei, Phys. Rev. D {\bf 73}, 083528 (2006); B. Hu, Y. Ling, Phys. Rev. D {\bf 73}, 123510 (2006).

\bibitem{obs} G. Olivares, F. Atrio-Barandela, D. Pavon, AIP Conf. Proc. {\bf 841}, 550 (2006); J-H. He, B. Wang, E. Abdalla, Phys. Rev. D {\bf 83}, 063515 (2011);
J-Q. Xia, Phys. Rev. D {\bf 80}, 103514 (2009).

\bibitem{zhou} J. Zhou, B. Wang, Y. Gong, E. Abdalla, Phys. Lett. B {\bf 652}, 86 (2007).

\bibitem{cao} S.A. Hayward, Class. Quant. Grav. {\bf 15}, 3147 (1998); R-G. Cai, L-M. Cao, Phys. Rev. D {\bf 75}, 064008 (2007)

\bibitem{aku}  M. Akbar, Int. J. Theor. Phys. {\bf 48}, 2665 (2009).

\bibitem{lima} J.A.S. Lima, J.S. Alcaniz, Phys. Lett. B {\bf 600}, 191 (2004); J. Zhou, B. Wang, D. Pav.on, E. Abdalla, Mod. Phys. Lett. A {\bf 24}, 1689 (2009).

\bibitem{kk} K. Karami, S. Ghaffari, Phys. Lett. B {\bf 685}, 115 (2010).

\bibitem{Bamba:2008hq}
  K.~Bamba, C.~Q.~Geng, S.~Nojiri and S.~D.~Odintsov,
  Phys.\ Rev.\  D {\bf 79}, 083014 (2009).

\bibitem{BGLL-BGL}
%
  K.~Bamba, C.~Q.~Geng, C.~C.~Lee and L.~W.~Luo,
  JCAP {\bf 1101}, 021 (2011);\ 
%
  K.~Bamba, C.~Q.~Geng and C.~C.~Lee,
  arXiv:1008.4036.
%

\bibitem{Jamil:2012fs} 
  M.~Jamil, D.~Momeni and R.~Myrzakulov,
  Eur.\ Phys.\ J.\ C {\bf 72}, 2137 (2012). 

\bibitem{Jamil:2012ju} 
  M.~Jamil, D.~Momeni and R.~Myrzakulov,
  Eur.\ Phys.\ J.\ C {\bf 72}, 2122 (2012). 

\bibitem{Jamil:2012vb} 
  M.~Jamil, D.~Momeni and R.~Myrzakulov,
  Eur.\ Phys.\ J.\ C {\bf 72}, 2075 (2012). 

\bibitem{Jamil:2012nm} 
  M.~Jamil, D.~Momeni, R.~Myrzakulov and P.~Rudra,
  J.\ Phys.\ Soc.\ Jap.\  {\bf 81}, 114004 (2012). 

\bibitem{Wei:2011yr} 
  H.~Wei,
  Phys.\ Lett.\ B {\bf 712}, 430 (2012). 

\bibitem{Xu:2012jf} 
  C.~Xu, E.~N.~Saridakis and G.~Leon,
  JCAP {\bf 1207}, 005 (2012). 

\bibitem{Gu:2012ww} 
  J.~-A.~Gu, C.~-C.~Lee and C.~-Q.~Geng,
  arXiv:1204.4048 [astro-ph.CO].

\bibitem{Bamba:2012mi} 
  K.~Bamba, C.~-Q.~Geng and L.~-W.~Luo,
  JCAP {\bf 1210}, 058 (2012). 

\bibitem{odintsov} S. Nojiri, S. D. Odintsov, S. Tsujikawa, Phys. Rev. D {\bf 71}, 063004 (2005).


\bibitem{Chen:2008sn} 
  F.~Chen, J.~M.~Cline and S.~Kanno,
  Phys.\ Rev.\ D {\bf 77}, 063531 (2008).

\end{thebibliography}
\end{document}